\documentclass[prl,aps,amsmath,amssymb,showpacs,twocolumn]{revtex4-1}

\usepackage{bm,color,float,dcolumn,amssymb,graphicx,hyperref,subfigure}

\begin{document}

\title{Fractional Topological Phases in Generalized Hofstadter Bands \\
with Arbitrary Chern Numbers}

\author{Ying-Hai Wu$^1$, J. K. Jain,$^1$ and Kai Sun$^2$}
\affiliation{$^1$ Department of Physics, The Pennsylvania State University, University Park, PA 16802, USA \\ 
$^2$ Department of Physics, University of Michigan, Ann Arbor, MI 48109, USA}

\date{\today}

\begin{abstract}
We construct generalized Hofstadter models that possess ``color-entangled" flat bands and study interacting many-body states in such bands. For a system with periodic boundary conditions and appropriate interactions, there exist gapped states at certain filling factors for which the ground state degeneracy depends on the number of unit cells along one particular direction. This puzzling observation can be understood intuitively by mapping our model to a single-layer or a multi-layer system for a given lattice configuration. We discuss the relation between these results and the previously proposed ``topological nematic states", in which lattice dislocations have non-Abelian braiding statistics. Our study also provides a systematic way of stabilizing various fractional topological states in $C>1$ flat bands and provides some hints on how to realize such states in experiments.
\end{abstract}

\maketitle

{\em Introduction} --- The topological structure of two-dimensional space plays an fundamental role in understanding the quantum Hall effect~\cite{Klitzing,Tsui}. It was proved by Thouless {\em et. al.}~\cite{Thouless} that, for a system of non-interacting electrons, the Hall conductance is proportional to the Chern number $C$ defined as the integral of Berry curvature over the Brillouin zone (BZ)~\cite{Simon}. This clarifies the topological origin of the integer quantum Hall effect because the Landau levels generated in an uniform magnetic field all have $C=1$. Haldane demonstrated subsequently that an uniform external magnetic field is not necessary by showing that a two-band model on honeycomb lattice with suitable parameters can have $C={\pm}1$ bands~\cite{Haldane1}; such systems are now termed ``Chern insulators"~\cite{Tang,Sun,Neupert}. When interactions between particles in a partially filled Landau level are taken into account, fractional quantum Hall (FQH) states can appear at certain filling factors. For a sufficiently flat band with a nonzero Chern number, fractional topological states may also be realized for suitable interactions~\cite{Sheng,Regnault,Wang1,Qi,Bernevig,Wang2,Zoo,Param,Goerbig,Murthy,Lee1,Venderbos,Cooper,Wu1,Repellin,Mine,Scaffidi,Liu,Lee2,Review1,Review2}. Many states in $C=1$ flat bands are shown to be adiabatically connected to those in Landau levels~\cite{Mine,Scaffidi,Liu}, which provides a simple way of characterizing their properties. 

In constrast to a Landau level which has $C=1$, a topological flat band can have an arbitrary Chern number. This motivates one to ask what is the nature of the fractional topological phases in $C>1$ flat bands~\cite{Wang3,Trescher,Bergholtz,Yang,Sterdyniak} and whether it is possible to realize some states that may not have analogs in conventional Landau levels. Ref.~\cite{Wu2} demonstrates that the incompressible ground states in a $C>1$ flat band at filling factor $\nu=1/(C+1)$ [$\nu=1/(2C+1)$] for bosons (fermions) can be interpreted as Halperin states~\cite{Halperin} with special flux insertions ({\em i.e.} boundary conditions) in {\em some} cases, but the nature of other states remains unclear. Ref.~\cite{Barkeshli1} proposed that some bilayer FQH states would have special properties if they are realized in $C=2$ systems and dubbed them as ``topological nematic states", but numerical evidence for such states has not been found. In this paper, we construct generalized Hofstadter models and demonstrate that there are interacting systems whose ground state degeneracy (GSD) depends on the number of unit cells along one direction. It is found that our models can be mapped either to a single-layer or to a multi-layer quantum Hall system depending on the lattice configuration, which provides a simple physical picture that helps us to understand the puzzling properties of $C>1$ flat bands. The change of GSD is one signature of the topological nematic states~\cite{Barkeshli1}, but there are other subtle issues, {\em e.g.} qualitative differences between bosons and fermions and the nature of the symmetry reduction, which we explain using our models.

\begin{figure}
\includegraphics[width=0.45\textwidth]{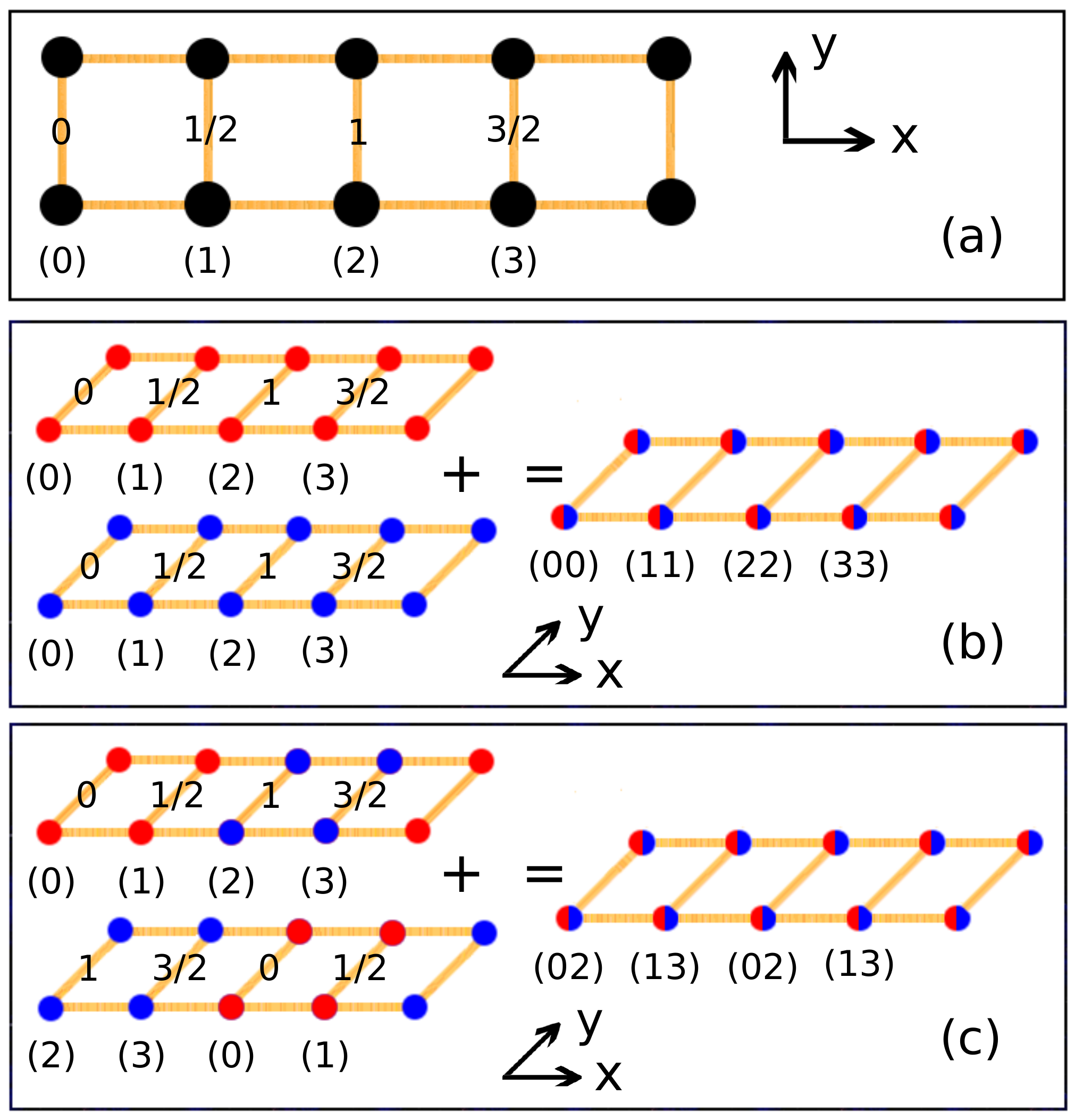}
\caption{(color online) In panel (a), we give an example of Hofstadter lattice with $n_\phi=4$ that are used in panels (b) and (c). The indices of orbitals in a magnetic unit cell are shown in parentheses and the numbers on the bonds indicate the phases of the complex hopping amplitudes along the $y$ direction in units of $\pi$. In panel (b), a bilayer Hofstadter model is obtained by stacking the two layers together. In panel (c), the two Hofstadter layers are shifted relative to each other and then stacked together. The method used in panel (c) gives a color-entangled Hofstadter model in which the size of the magnetic unit cell is reduced by a factor of two and the lowest band has $C=2$. There are two orbitals on each lattice site (colored in red and blue) for both models and their indices are given in parentheses.}
\label{Figure1}
\end{figure}

{\em Color-Entangled Hofstadter Models} --- We construct flat bands with arbitrary Chern numbers by generalizing the Hofstadter model~\cite{Peierls,Harper,Wannier,Azbel,Hofstadter} using the generic scheme of Ref.~\cite{Yang}. As shown in Fig~\ref{Figure1} (a), the Hofstadter model describes particles in both a uniform magnetic field and a periodic potential. The tight-binding Hamiltonian for the model on a square lattice is $H=\sum_{ij} t_{ij} e^{i\theta_{ij}} {\widehat a}^\dagger_i {\widehat a}_j + {\rm H.c.}$, where $\theta_{ij}$ is the phase associated with the hopping from site $j$ to $i$~\cite{Peierls}, ${\widehat a}^\dagger_i$ is the creation operator on site $i$ and ${\rm H.c.}$ means Hermitian conjugate. If the magnetic flux per plaquette is $2\pi/n_{\phi}$ with $n_{\phi}$ being an integer, translational symmetry is preserved on the scale of magnetic unit cell which contains $n_\phi$ plaquettes. The momentum space Hamiltonian is $H(\mathbf k)=\Psi^\dagger(\mathbf k)\mathcal{H}\Psi(\mathbf k)$, where $\Psi^\dagger(\mathbf k)=[{\widehat a}^\dagger_0(\mathbf k), {\widehat a}^\dagger_1(\mathbf k), \cdots, {\widehat a}^\dagger_{n_{\phi}-1}(\mathbf k)]$ and the subscript of ${\widehat a}^\dagger$ marks different sites within a magnetic unit cell. $\mathcal{H}(\mathbf k)$ is a $n_{\phi}{\times}n_{\phi}$ matrix whose non-zero matrix elements are $\mathcal{H}_{mm}(\mathbf k) = 2\cos(k_y+2m\pi/n_\phi)$ and $\mathcal{H}_{mn}(\mathbf k)=\mathcal{H}^*_{nm}(\mathbf k) = \exp(ik_x/n_\phi)$ for $n=(m+1) \; {\rm mod} \; n_\phi$. The $n_\phi\rightarrow \infty$ limit recovers the continuous limit in which Landau levels arise. One important advantage of starting from the Hofstadter model is that the Berry curvature of the lowest band can be made uniform over the entire BZ. This is desirable because a nonuniformity in the Berry curvature usually tends to weaken or even destroy incompressible states~\cite{Mine}. 

We next stack two identical Hofstadter lattices together in two different ways. In Fig.~\ref{Figure1} (b), the same orbitals in different layers are aligned together, which results in a conventional bilayer quantum Hall system. In Fig.~\ref{Figure1} (c), the $m$-th orbital in one layer is aligned with the $(m+n_\phi/2)$-th orbital in the other layer. The latter stacking pattern reduces the size of the magnetic unit cell by half~\cite{Yang}, so the model shown in Fig~\ref{Figure1} (c) possess a single lowest band with $C=2$ instead of having two degenerate $C=1$ bands. It should be emphasized that these two systems are equivalent insofar as the behavior of the {\em bulk} is concerned, since they correspond to two different gauge choices. However, as shown below, they behave differently when periodic boundary conditions (PBCs) are imposed because PBCs are not invariant under a change of gauge. In general, one can get a band with an arbitrary Chern number $C$ by stacking $C$ layers of Hofstadter lattices and aligning the orbitals labeled by $m$, $m+n_\phi/C$, $\cdots$, and $m+(C-1)n_\phi/C$ in different layers (where $m\in[0,1,\cdots,n_\phi/C-1]$). The momentum space Hamiltonian is very similar to the single-layer Hofstadter model, except that the off-diagonal term $\mathcal{H}_{mn}(\mathbf k)$ is replaced by $\exp(ik_xC/n_\phi)$. A detailed comparison revealing the similarity of the ``color-entangled Bloch basis"~\cite{Wu2} and our models is given in Appendix A, which suggests that our models, as well as those constructed in Ref.~\cite{Yang}, can be referred to as ``color-entangled" topological flat band models. 

{\em Interacting Many Body Systems} --- The differences between a bilayer Hofstadter model and a $C=2$ color-entangled Hofstadter model become transparent when one studies interacting many body systems. We consider $N$ particles on a periodic lattice with $N_x$ and $N_y$ magnetic unit cells along the $x$ and $y$ directions. The total number of plaquettes (to be distinguished with the numbers of magnetic unit cells) along the $x$ and $y$ directions are denoted as $L_x$ and $L_y$, respectively. The magnetic unit cell is always chosen to contain only one plaquette in the $y$ direction so we have $L_y=N_y$. It was proposed in Ref.~\cite{Barkeshli1} that the following bilayer quantum Hall wave functions
\begin{eqnarray}
\Psi_B(\{z^1\},\{z^2\}) &=& \Phi_{\frac{p}{p+1}}(\{z^1\}) \Phi_{\frac{p}{p+1}}(\{z^2\}) 
\label{BoseMultiWave} \\
\Psi_F(\{z^1\},\{z^2\}) &=& \prod_{i<j} (z^1_i-z^1_j)^3 \prod_{i<j} (z^2_i-z^2_j)^3 \nonumber \\
&\times& \prod_{i,j} (z^1_i-z^2_j) 
\label{FermiMultiWave}
\end{eqnarray}
are topological nematic states in $C=2$ bands, where $z=x+iy$ is the complex coordinate and its superscript indicates the layer it resides in. Eq.~(\ref{BoseMultiWave}) describes bosonic systems with two decoupled layers and Eq.~(\ref{FermiMultiWave}) is the Halperin $331$ state~\cite{Halperin} for fermions. The value of $p$ is chosen to be $1$ or $2$ and the associated wave functions are the Laughlin $1/2$ state $\Phi_{1/2}(\{z^\alpha\})$ and the Jain $2/3$ state $\Phi_{2/3}(\{z^\alpha\})$ (they are the bosonic analogs of the Laughlin $1/3$ state~\cite{Laughlin} and the Jain $2/5$ state~\cite{Jain} for fermions). When the states represented by Eq.~(\ref{BoseMultiWave}) and Eq.~(\ref{FermiMultiWave}) are realized on a torus with PBCs, the GSDs are $(p+1)^2$ and $8$ respectively~\cite{Haldane2,McDonald}.

\begin{figure}
\includegraphics[width=0.45\textwidth]{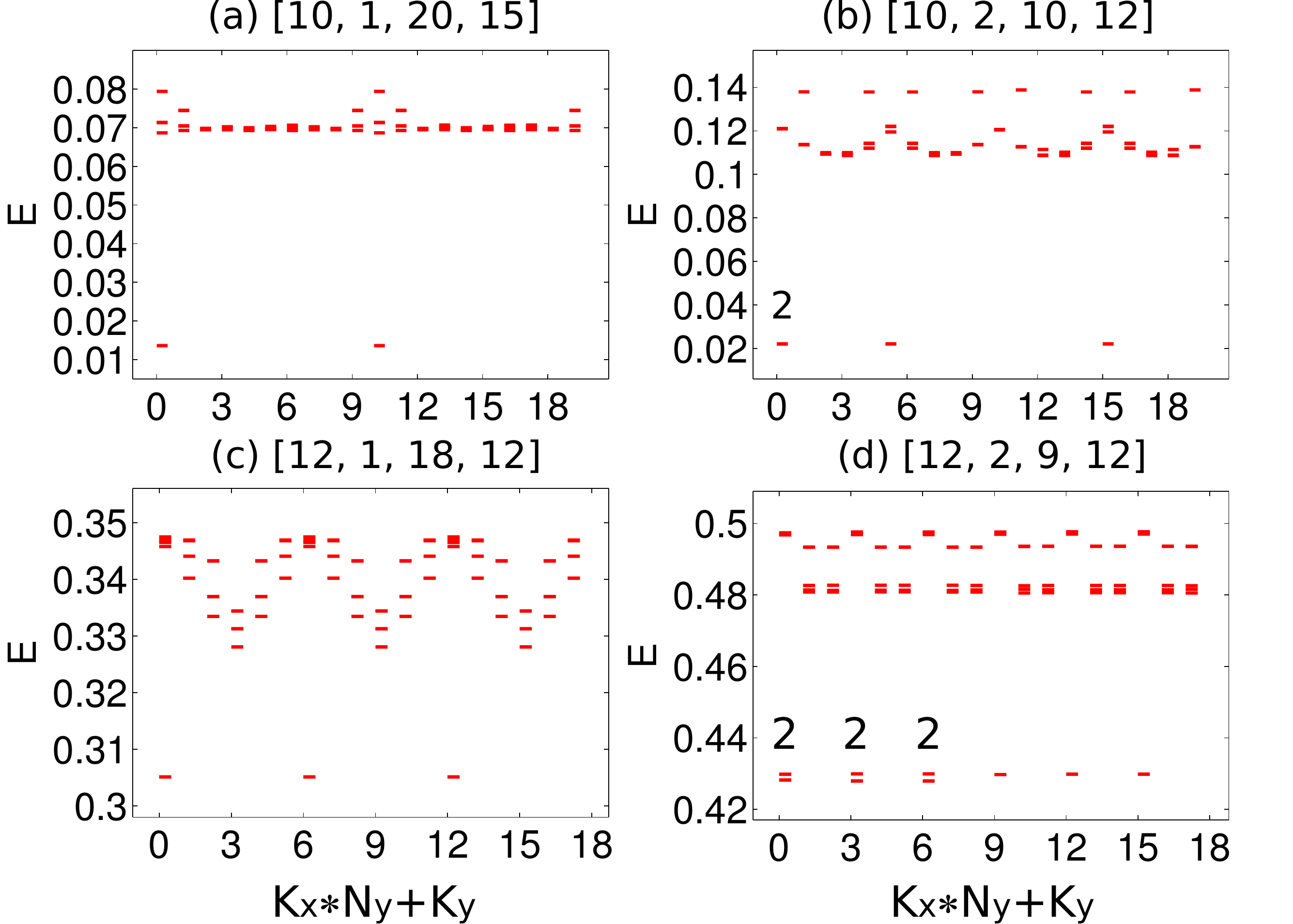}
\caption{Energy spectra of bosons on the $C=2$ model at filling factors $1/2$ [(a) and (b)] and $2/3$ [(c) and (d)]. The system parameters are given in square brackets as $[N, N_x, N_y, L_x]$. The numbers above some energy levels indicate degeneracies that may not be resolved by inspection.}
\label{Figure2}
\end{figure}

\begin{figure}
\includegraphics[width=0.45\textwidth]{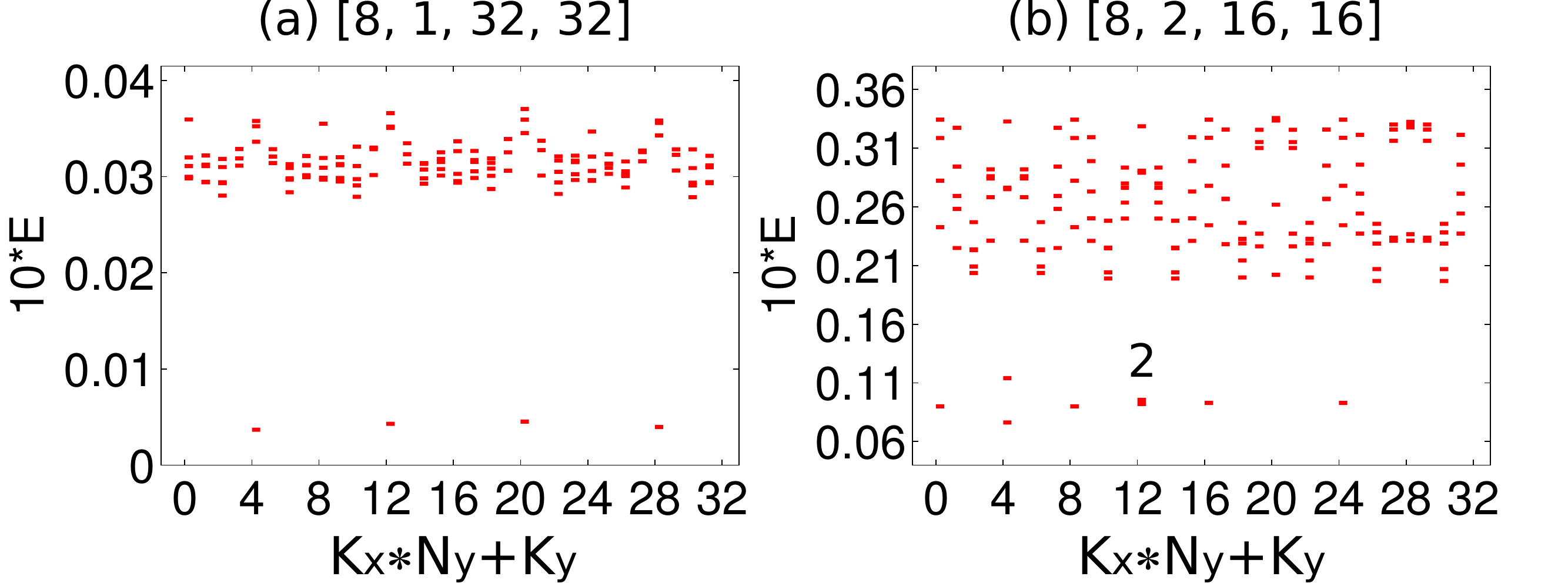}
\caption{Energy spectra of fermions on the $C=2$ model at filling factor $1/4$. The system parameters are given in square brackets as $[N, N_x, N_y, L_x]$. The numbers above some energy levels indicate degeneracies that may not be resolved by inspection.}
\label{Figure3}
\end{figure}

The wave functions Eq.~(\ref{BoseMultiWave}) and Eq.~(\ref{FermiMultiWave}) can be realized in a bilayer Landau level system in continuum when intra-layer interaction is stronger than inter-layer interaction. This motivates us to study the Hamiltonians $H_B = \sum_i \sum_{\sigma} U^B_{\sigma\sigma} : {\widehat n}_i(\sigma) {\widehat n}_i(\sigma) : + \sum_i \sum_{\sigma\neq\tau} V^B_{\sigma\tau} : {\widehat n}_i(\sigma) {\widehat n}_i(\tau) : $ and $H_F=\sum_{\langle{ij}\rangle} \sum_\sigma U^F_{\sigma\sigma} : {\widehat n}_i(\sigma) {\widehat n}_j(\sigma) : + \sum_{i} \sum_{\sigma{\neq}\tau} V^F_{\sigma\tau} : {\widehat n}_i(\sigma) {\widehat n}_i(\tau) :$, where $:\cdots:$ enforces normal ordering, ${\widehat n}_i(\sigma)$ is the number operator for particle of color $\sigma$ on site $i$, and $\langle{ij}\rangle$ denotes nearest neighbors. The parameters are chosen as $U^B_{\sigma\sigma}=1$, $V^B_{\sigma\tau}=0.03$, $U^F_{\sigma\sigma}=1$ and $V^F_{\sigma\tau}=0.5$. In other words, we use intra-color onsite interactions for bosonic systems (with a small perturbation given by the $V^B_{\sigma\tau}$ terms~\cite{Note1}) and both intra-color NN and inter-color onsite interactions for fermionic systems. The eigenstates of these Hamiltonians are labeled by their momenta $K_x$ and $K_y$ along the two directions. The many-body Hamiltonians are projected into the partially occupied lowest band(s)~\cite{Regnault} and the filling factor is defined as $\nu=N/(\mathcal{M}N_xN_y)$, where $\mathcal{M}$ is the number of bands that are kept in the projection ({\em i.e.}, $\mathcal{M}=2$ for the bilayer Hofstadter model and $\mathcal{M}=1$ for the $C=2$ color-entangled Hofstadter model). This means that Eq.~(\ref{BoseMultiWave}) and Eq.~(\ref{FermiMultiWave}) have filling factors $p/(p+1)$ and $1/4$ respectively. 

The number of plaquettes in the $x$ direction for given $N_x$ and $N_y$ values is chosen to ensure that this system is close to isotropic ({\em i.e.} has aspect ratio close to 1). As the size of the unit cell increases, the wave function of a particle spreads over a larger area and the interaction between two particles becomes weaker. To compare systems with different $n_{\phi}$, we normalize the energy scale using the total energy of two particles in a system with $N_x=1$ and $N_y=1$ ($N_x=1$ and $N_y=2$) for bosons (fermions).

The GSDs of Eq.~(\ref{BoseMultiWave}) and Eq.~(\ref{FermiMultiWave}) given above were derived using the continuum wave functions, but we have found that they are still valid for bosonic systems with Hamiltonian $H_B$ and fermionic systems with Hamiltonain $H_F$ in the bilayer Hofstadter model. The results in the $C=2$ color-entangled Hofstadter model are very different as presented in Fig.~\ref{Figure2} and Fig.~\ref{Figure3}. The special feature of the $C=2$ systems is that the GSD depends on $N_x$ and it only agrees with the result in the bilayer Hofstadter model when $N_x$ is even. For bosonic systems, the GSD at $1/2$ is $2$ if $N_x$ is odd and $4$ if $N_x$ is even; the GSD at $2/3$ is $3$ if $N_x$ is odd and $9$ if $N_x$ is even. The gaps of the bosonic states survive in the presence of small inter-color onsite interaction $V^B_{\sigma\tau}$ but disappear if $V^B_{\sigma\tau}$ becomes comparable to $U^B_{\sigma\tau}$. The phase boundary can not be determined precisely because a reliable finite-size scaling is difficult here. For fermionic systems, the GSD is $4$ for $N_x=1$ and $8$ for $N_x=2$. The quasi-degenerate ground states have a more prounced splitting than the bosonic cases. The gaps become less clear for larger $N_x$ and there is no well-defined set of quasi-degenerate ground states when $N=8$, $N_x=4$, $N_y=8$, and $L_x=16$.

\begin{figure}
\includegraphics[width=0.45\textwidth]{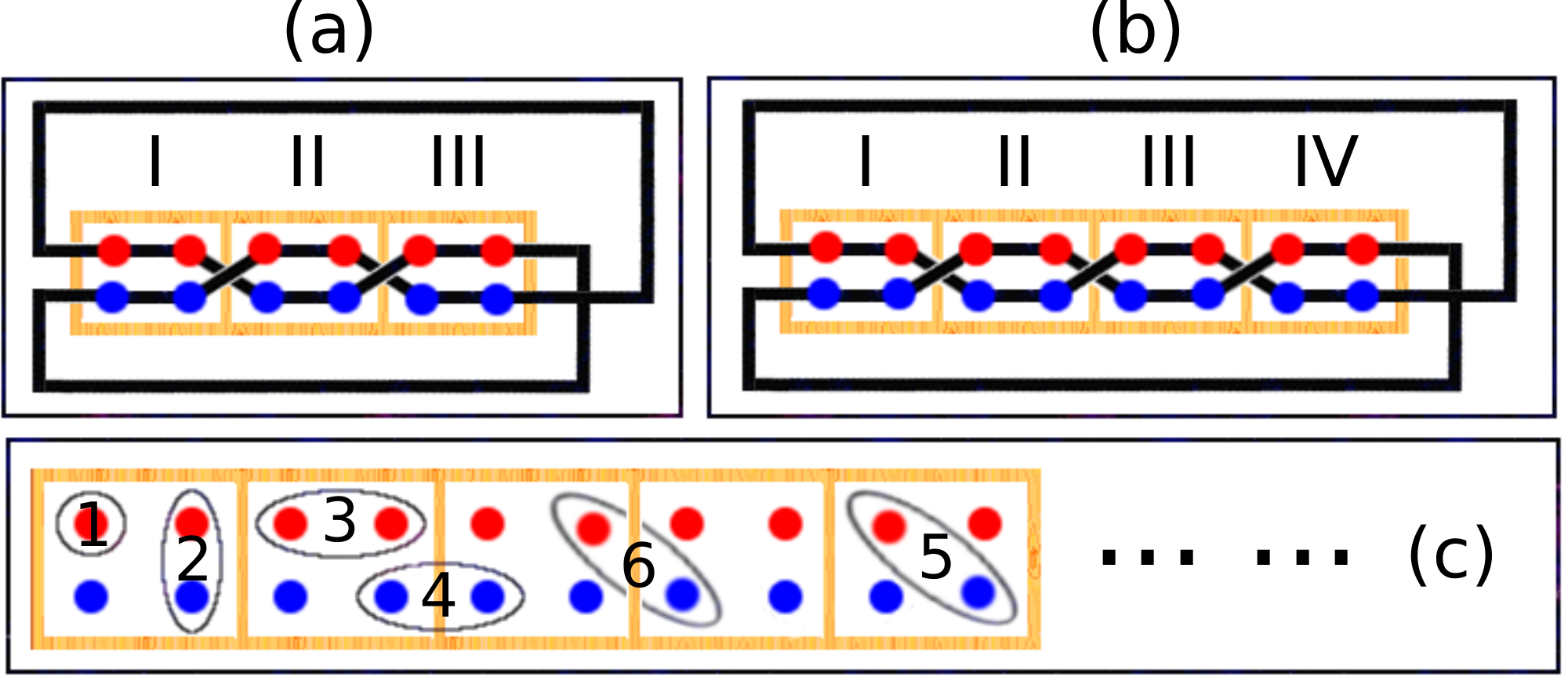}
\caption{(color online) This figure shows a slice of the $C=2$ model constructed in Fig.~\ref{Figure1} but the two orbitals are plotted separately for clarity. In panels (a) and (b), the unit cells are labeled by Roman numbers and the black lines represent the hopping terms along the $x$ direction. When $N_x$ is odd in (a) [even in (b)], this model maps into a single-layer (bilayer) system. The hopping terms along the $y$ direction do not change this mapping. Panel (c) shows certain interaction terms: 1. intra-color onsite term; 2. inter-color onsite term; 3. intra-color NN term within one unit cell; 4. intra-color NN term across the boundary of a unit cell; 5. inter-color NN term within one unit cell; 6. inter-color NN term across the boundary of a unit cell.}
\label{Figure4}
\end{figure}

{\em Boundary Conditions, Topology, and Symmetry} --- The key to understanding the physics of a $C=2$ band is that it may have two fundamentally distinct topologies determined by the parity of $N_x$. As illustrated in Fig.~\ref{Figure4}, this originates from the twisted hoppings along the $x$ direction at the boundary between two magnetic unit cells ({\em i.e.} the color index of a particle is flipped). For odd $N_x$ [Fig.~\ref{Figure4} (a)], it can be unfolded to produce a single Hofstadter layer with $C=1$ by tracking the black lines which represent hopping terms along the $x$ direction. For even $N_x$ [Fig.~\ref{Figure4} (b)], it contains two decoupled Hofstadter layers each having $C=1$. This mapping is sufficient to explain why the GSD change in the bosonic systems: a single-layer $p/(p+1)$ state with GSD $p+1$ is realized for odd $N_x$, while two decoupled $p/(p+1)$ states with GSD $(p+1)^2$ appear when $N_x$ is even. The fermionic case is more complicated, but it was argued that the GSD of the Halperin $331$ state in a $C=2$ band is $8$ when $N_x$ is even and $4$ if $N_x$ is odd~\cite{Barkeshli1}. 

The fact that the unfolding of the model depends only on the parity of $N_x$ but not of $N_y$ signifies a reduction of rotational symmetry. However, in our systems the $C_4$ symmetry is not broken spontaneously, as in previously studied nematic states~\cite{Kivelson,Fradkin}, but results from the model Hamiltonian itself through boundary conditions. To gain insight into this issue, we note that the simple square lattice Hofstadter model has four-fold rotational symmetry $C_4$ (up to gauge transformations) even though a magnetic unit cell usually has only two-fold rotational symmetry $C_2$. This conclusion is valid when the system contains an integral number of magnetic unit cells, which is also satisfied automatically for a Hofstadter mutli-layer. In contrast, since the unit cell of the color-entangled $C=2$ Hofstadter model is half as large as the original magnetic unit cell, the $C_4$ symmetry of the parent Hofstadter model is inherited only when $N_x$ is even, but is reduced to $C_2$ symmetry for odd $N_x$. 

Although the GSDs of bosonic systems confirm the theoretical predictions, the fermionic $331$ state seems less stable. This puzzle is resolved when we analyze the 2-body interaction terms shown in Fig.~\ref{Figure4} (c), which also have different effects depending on the parity of $N_x$. For both even and odd $N_x$, the term (1) in Fig.~\ref{Figure4} (c) is still an onsite term and both (3) and (6) turn out to be intra-layer nearest neighbor (NN) terms. If $N_x$ is even, (2), (4) and (5) become, respectively, an inter-layer onsite term, an inter-layer NN term, and an inter-layer NN term. On the other hand, when $N_x$ is odd, (2), (4) and (5) all result in interactions, in the single unfolded layer, that extend over a range comparable to the system size. In a fermionic system with even $N_x$, the intra-color NN terms across boundaries between unit cells turn into inter-layer NN terms when the model is mapped to a bilayer system, which is expected to weaken or destroy the $331$ state. If one carefully design the Hamiltonian to make sure that it contains no inter-layer NN terms after mapping into a bilayer system, then one can get a $331$ state with a clear gap~\cite{Note2}. In Appendix B, we discuss the nature of the gapped ground states observed in previous works~\cite{Wang2,Trescher,Bergholtz,Yang,Sterdyniak} in light of these observations. In Appendix C, we discuss a subtle issue in the square lattice $C=2$ model~\cite{Yang} and explain how to observe the change of GSD in this model.

How relevant is the above analysis using PBCs for real physical systems with open boundaries? The topology of a lattice is determined by $N_x$ because the hoppings along the $x$ direction at each boundary between two magnetic unit cells flip the color indices of particles. It was proposed that edge dislocations have a similar effect~\cite{Barkeshli1} and they have {\em projective} non-Abelian braiding statistics~\cite{Barkeshli2}: there are multiple degenerate states given a fixed configuration of dislocations; an exchange of two dislocations results in a unitary evolution in this degenerate space; two such exchanges may not commute with each other; the overall phases of the braiding operations are undetermined. These exotic properties may be demonstrated in tunneling and interferometric measurements~\cite{Barkeshli3}. The physics of defects in various topological phases have also been studied~\cite{Cheng,Lindner,Clarke,Vaezi}.

We finally briefly mention possible experimental realizations of the fractional topological phases studied above. A variety of proposals for creating synthetic gauge fields for cold atoms have been studied~\cite{Dalibard,Goldman1} and the complex hoppings in the standard Hofstadter model have been successfully implemented~\cite{Aidelsburger,Miyake}. The feasibility of simulating topological phases using photons has also been investigated~\cite{Koch1,Umucalilar1,Hafezi1,Houck,Umucalilar2,Kapit1,Hafezi2,Kapit2}. The standard Hofstadter model and its time-reversal symmetric variant have been demonstrated for non-interacting photons~\cite{Hafezi3,Rechtsman,Jia,Mittal}. In view of these achievements and following similar lines of thought, we discuss how color-entangled Hofstadter models with $C>1$ bands may be realized in Appendix D.

In conclusion, we have constructed color-entangled Hofstadter models with arbitrary Chern numbers and demonstrated the existence of fractional topological phases in such systems by extensive exact diagonalization studies. The models we use help to clarify many aspects of topological flat bands with $C>1$ in a physically intuitive manner.

{\em Acknowledgements} --- Numerical calculations are performed using the DiagHam package, for which we are very grateful to the authors, especially N. Regnault and Y.-L. Wu. YHW and JKJ thank V. B. Shenoy for many helpful discussions on the possible experimental realizations and the Indian Institute of Science for their kind hospitality. YHW and JKJ are supported by DOE under Grant No. DE-SC0005042. KS is supported in part by NSF under Grant No. ECCS-1307744 and the MCubed program at University of Michigan. High-performance computing resources and services are provided by Research Computing and Cyberinfrastructure, a unit of Information Technology Services at The Pennsylvania State University.

\clearpage

\setcounter{figure}{0}
\setcounter{equation}{0}
\renewcommand\thefigure{S\arabic{figure}}
\renewcommand\theequation{S\arabic{equation}}

\begin{appendix}

\begin{widetext}

\subsection{Appendix A. Color-Entangled Bloch Basis and Hofstadter Model}

We first briefly review the color-entangled Bloch basis for Landau levels~\cite{Wu2}. The unit vectors along the $x$ and $y$ directions are denoted as ${\widehat e}_x$ and ${\widehat e}_y$. We consider a torus defined by vectors ${\mathbf L}_1=L_1{\widehat e}_v$ and ${\mathbf L}_2=L_2{\widehat e}_y$ with ${\widehat e}_v = \sin\theta {\widehat e}_x + \cos\theta {\widehat e}_y$. The magnetic field $B$ along the $z$ direction is generated by the Landau gauge vector potential ${\mathbf A}({\mathbf r})=Bx{\widehat e}_y$. The magnetic translation operator is $T({\mathbf a})=e^{-i{\mathbf K}\cdot{\mathbf a}}$ with ${\mathbf K}=-i\hbar\nabla-e{\mathbf A}+e{\mathbf B}\times{\mathbf r}$. To satisfy periodic boundary conditions $T({\mathbf L}_\alpha)=1$, the number of magnetic flux through the torus has to be an integer $N_\phi=L_1L_2\sin\theta/(2{\pi}\ell^2_B)$. The lowest Landau level wave functions have the form
\begin{eqnarray}
\langle {\mathbf r}|j\rangle = \frac{1}{(\sqrt{\pi} L_2 \ell_B)^{1/2}} \sum_n^{\mathbb{Z}} \exp\Big[ 2\pi(j+nN_\phi)\frac{x+iy}{L_2} -i\frac{\pi L_1 e^{-i\theta}} {N_\phi L_2} (j+nN_\phi)^2 \Big] e^{-x^2/(2\ell^2_B)}
\end{eqnarray}
with index $j\in[0,1,\cdots,N_\phi-1]$. For a pair of integers $N_x$ and $N_y$ satisfying $N_\phi=N_xN_y$, we can construct Bloch basis states
\begin{eqnarray}
|k_x,k_y\rangle &=& \frac{1}{\sqrt{N_x}}\sum_{m=0}^{N_x-1}e^{i2\pi mk_x/N_x} |j=mN_y+k_y\rangle
\end{eqnarray}
with indices $k_x\in[0,1,\cdots,N_x-1]$ and $k_y\in[0,1,\cdots,N_y-1]$, which are eigenstates of translation operators $T_x=T(\mathbf{L}_1/N_x)$ and $T_y=T(\mathbf{L}_2/N_y)$.

To construct basis states for multi-component Landau levels, whose internal degree of freedom is dubbed as color, we introduce two color operators $P$ and $Q$ that act on a color eigenstate $|s\rangle$ as follows
\begin{equation}
P|s\rangle = |s+1\text{ (mod $C$)}\rangle \;\;\; Q|s\rangle = e^{i2{\pi}s/C}|s\rangle
\end{equation}
This means that $P$ flips the color index and $Q$ imprints a phase according to the color index. We entangle translation in real space and rotation in the internal color space using two commuting operators $\widetilde{T}_x=T_xP$ and $\widetilde{T}_y=T_yQ$. The basis states can be chosen as
\begin{eqnarray}
\langle {\mathbf r},s|k_x,k_y\rangle &=& \frac{1}{(\sqrt{\pi} N_x L_2 \ell_B)^{1/2}} \sum_n^{\mathbb{Z}} e^{i2\pi(nC+s)k_x/N_x} e^{-x^2/(2\ell^2_B)} 
\nonumber \\ 
&\times& \exp\Bigg[ 2\pi\left(k_y+nN_y+\frac{s}{C}N_y\right)\frac{x+iy}{L_2} - i\frac{{\pi} L_1 e^{-i\theta}}{N_\phi L_2}\left(k_y+nN_y+\frac{s}{C}N_y\right)^2 \Bigg]
\end{eqnarray}
with $N_\phi=N_xN_y/C$ and $s\in[0,1,\cdots,C-1]$, which satisfy ${\widetilde T}_\alpha|k_x,k_y\rangle=\exp(-i2\pi{k_\alpha}/N_\alpha)|k_x,k_y\rangle$. The color-entangled boundary conditions ${\widetilde{T}_\alpha}^{N_\alpha}=1$ impose the constraints $k_\alpha\in\mathbb{Z}$ so the primiary region can be chosen as $k_x\in[0,1,\cdots,N_x-1]$ and $k_y\in[0,1,\cdots,N_y-1]$. 

We now demonstrate that the Bloch basis discussed above is closely related to the generalized Hofstadter models using the four band $C=2$ model shown in Fig.~\ref{Figure1} as an example. The moment space single-particle Hamiltonian is
\begin{eqnarray}
H({\mathbf k}) =
\Psi^\dagger({\mathbf k}) \left(
\begin{array}{cccc}
2\cos(k_y) & e^{ik_x/2} & 0 & e^{-ik_x/2} \\
e^{-ik_x/2} & 2\cos(k_y+\pi/2) & e^{ik_x/2} & 0 \\
0 & e^{-ik_x/2} & 2\cos(k_y+\pi) & e^{ik_x/2} \\
e^{ik_x/2} & 0 & e^{-ik_x/2} & 2\cos(k_y+3\pi/2) \\
\end{array}
\right) \Psi({\mathbf k})
\label{Hamilton1}
\end{eqnarray}
where $\Psi^\dagger({\mathbf k})=[a^\dagger_0({\mathbf k}),a^\dagger_1({\mathbf k}),a^\dagger_2({\mathbf k}),a^\dagger_3({\mathbf k})]$. It can be rewritten as
\begin{eqnarray}
H({\mathbf k}) &=& \Psi^\dagger_{02}({\mathbf k}) e^{ik_y} \left(
\begin{array}{cc}
1 & 0 \\
0 & e^{i\pi} \\
\end{array}
\right) \Psi_{02}({\mathbf k}) + \Psi^\dagger_{13}({\mathbf k}) e^{ik_y} \left(
\begin{array}{cc}
e^{i\pi/2} & 0 \\
0 & e^{i3\pi/2} \\
\end{array}
\right) \Psi_{13}({\mathbf k}) \nonumber \\
&+& \Psi^\dagger_{02}({\mathbf k}) e^{ik_x/2} \left(
\begin{array}{cc}
1 & 0 \\
0 & 1 \\
\end{array}
\right) \Psi_{13}({\mathbf k}) + \Psi^\dagger_{13}({\mathbf k}) e^{ik_x/2} \left(
\begin{array}{cc}
0 & 1 \\
1 & 0 \\
\end{array}
\right) \Psi_{02}({\mathbf k}) + {\rm H. \; c.}
\label{Hamilton2}
\end{eqnarray}
where $\Psi^\dagger_{02}=(a^\dagger_0,a^\dagger_2)$ and $\Psi^\dagger_{13}=(a^\dagger_1,a^\dagger_3)$. On the right hand side of Eq.~(\ref{Hamilton2}), the first two terms describe hopping in the $y$ direction, where the second component always has an additional $\pi$ phase relative to the first; the third term originates from hopping in the $x$ direction {\em within} a unit cell; the fourth term comes from hopping in the $x$ direction {\em across} a boundary separating adjacent unit cells. The color index is flipped when the particle hops across the boundary, as the hopping matrix connecting $\Psi^\dagger_{13}$ and $\Psi_{02}$ is off-diagonal. These effects are the same as the color-entangled boundary conditions discussed above. 

\subsection{Appendix B. Ground States at $\nu=1/(C+1)$ and $\nu=1/(2C+1)$}

For certain color-independent Hamiltonians and the $C$-color Bloch basis, zero energy ground state occur at filling factor $\nu=1/(C+1)$ [$\nu=1/(C+1)$] for bosons (fermions)~\cite{Wu2}. These states correspond to color-dependent flux inserted version of the Halperin states~\cite{Halperin,Wu2} when $N_x$ is a multiple of $C$ ({\em i.e.} in the cases where the Bloch basis can be mapped to a multi-layer system). We have confirmed that these FQH states also appear in  our color-entangled Hofstadter models with Chern number $C$ by using the Hamiltonian
\begin{eqnarray}
{\widetilde H}_{\rm B} = \sum_i \sum_{\sigma} : n_i(\sigma) n_i(\sigma) : + \sum_i \sum_{\sigma\neq\tau} : n_i(\sigma) n_i(\tau) :
\label{BoseHamilton}
\end{eqnarray}
for bosons and the Hamiltonian 
\begin{eqnarray}
{\widetilde H}_{\rm F} &=& \sum_{\langle{ij}\rangle} \sum_\sigma : n_i(\sigma) n_j(\sigma) : + \sum_{i} \sum_{\sigma{\neq}\tau} : n_i(\sigma) n_i(\tau) : + \sum_{\langle{ij}\rangle} \sum_{\sigma{\neq}\tau} : n_i(\sigma) n_j(\tau) :
\label{FermiHamilton}
\end{eqnarray}
for fermions. The energy spectra as well as particle entanglement spectra match the results obtained using color-entangled Bloch basis.

Based on our analysis in the main text and the previous section, the $C$-color Hofstadter model or Bloch basis can be mapped to a single layer if $N_x$ is not a multiple of $C$, but the nature of the gapped states here is unclear. As shown in Fig.~\ref{Figure2}, some local interaction terms in Eq.~(\ref{BoseHamilton}) and Eq.~(\ref{FermiHamilton}) induce special long-range correlations when the system is unfolded to a single layer, but their exact forms in the continnum are not known without analytical calculations. To test this interpretation more explicitly, we have tested many different Hamiltonians for particles in a one-component Landau level on torus and found that {\em some} choices of system-dependent unnatural long-range interactions (in addition to short-range ones) indeed produce gapped ground states at filling factor $1/3$ ($1/5$) for bosons (fermions). 

\begin{figure}
\includegraphics[width=0.6\textwidth]{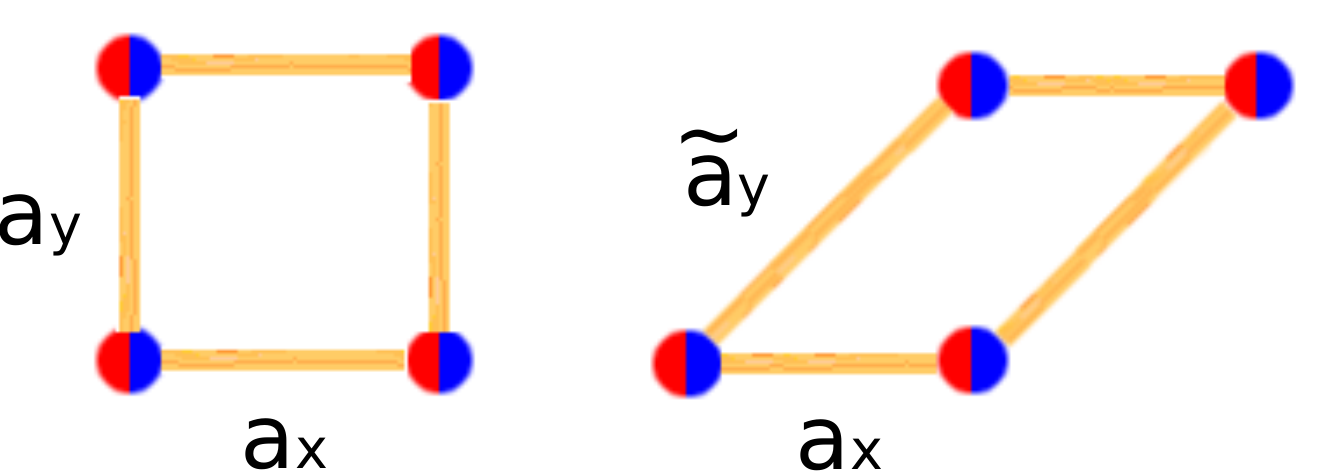}
\caption{(color online) Square lattice two-orbital model with Chern number $C=2$. The red and blue colors on each site represent the two orbitals.}
\label{FigureS1}
\end{figure}

\begin{figure}
\includegraphics[width=0.6\textwidth]{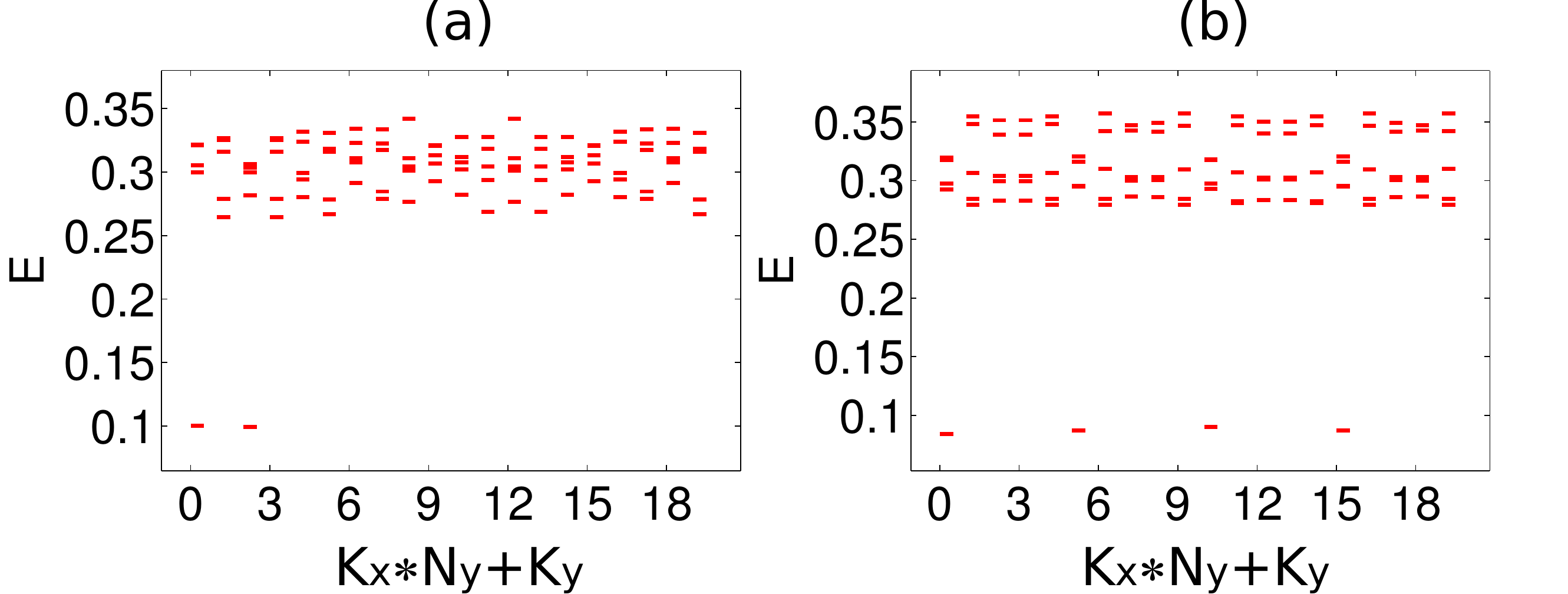}
\caption{Ground state energy spectra of bosons on the square lattice $C=2$ model at filling factor $\nu=1/2$. (a) $N=10$, $N_x=5$, $N_y=4$; (b) $N=10$, $N_x=4$, $N_y=5$.}
\label{FigureS2}
\end{figure}

\subsection{Appendix C. Square Lattice $C=2$ Model}

Here we generalize our considerations to the square lattice $C=2$ model~\cite{Yang}, which can be obtained by stacking two checkerboard lattices together and shift them relative to each other along the ${\mathbf a}_x$ direction defined in Fig.~\ref{FigureS1}. The checkerboard lattice model~\cite{Sun} contains two orbitals per unit cell and there are NN, next NN and second next NN hopping terms. The NN hopping terms connect the two types of orbitals and this brings out certain additional subtleties. For the usual choice of the primitive translation vectors ${\mathbf a}_x$ and ${\mathbf a}_y$ (Fig.~\ref{FigureS1}), the hoppings along both these directions are associated with changes of color index. A system is mapped to a single layer when one of $N_x$ and $N_y$ is odd, and two decoupled layers when both are even. Insight into the physics of this model is given by choosing instead ${\mathbf a}_x$ and ${\widetilde{\mathbf a}}_y$ to define the unit cell. In this case, the color index of a particle does not change during hopping along the ${\widetilde{\mathbf a}}_y$ direction, and a system may be mapped to a single layer or two layers depending {\em only} on the parity of $N_x$. The momentum space single-particle Hamiltonian in this case is
\begin{eqnarray}
\mathcal{H}_{\rm S} &=& 2t_3 \left[ \cos(2k_x) + \cos(2k_y-2k_x) \right] \mathbb{I} + \sqrt{2} t_1 \left[ \cos(k_x) + \cos(k_y-k_x) \right] \sigma_x \nonumber \\
  &-& \sqrt{2} t_1 \left[ \cos(k_x) - \cos(k_y-k_x) \right] \sigma_y - 4 t_2 \sin(k_x)\sin(k_y-k_x) \sigma_z 
\end{eqnarray}
where $t_1=1$, $t_2=1/(2+\sqrt{2})$ and $t_3=1/(2\sqrt{2}+2)$. $\mathbb{I}$ is the identity matrix and $\sigma_{x,y,z}$ are the Pauli matrices. This Hamiltonian is different from the one given in Ref.~\cite{Yang}, because we are using different lattice translation vectors. We use 2-body onsite interaction given by $H_{\rm S}=\sum_i : n_i(A) n_i(A) + n_i(B) n_i(B) + 0.06 n_i(A) n_i(B) :$ (the small interaction between $A$ and $B$ is used to split some degeneracies and increase the speed of exact diagonalization). The energy spectra of bosons on this model are shown in Fig.~\ref{FigureS2}. We see that the GSD is $2$ when $N_x=5$ and $4$ when $N_x=4$, which can be understood along the same lines as for the models discussed in the main text.

\begin{figure}
\includegraphics[width=0.9\textwidth]{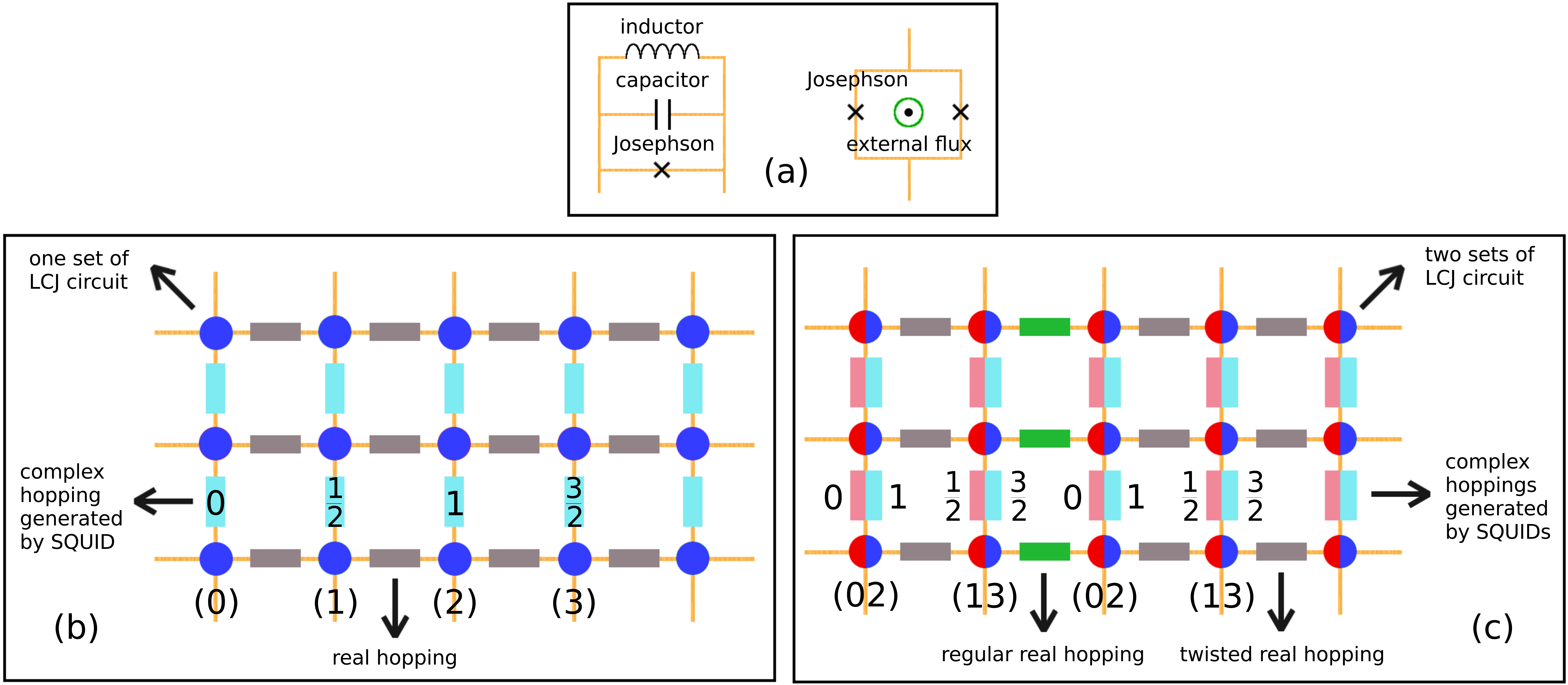}
\caption{Schematics of the standard Hofstadter model and the color-entangled Hofstadter model implemented using circuit quantum electrodynamics components. The basic elements of the systems are shown in panel (a). The left part is an LCJ circuit consists of one inductor, one capacitor, and one Josephson junction. The right part is a SQUID made of two Josephson junctions and controlled by an external flux. The standard Hofstadter model can be realized using the circuit shown in panel (b). There is one LCJ circuit on each lattice site and they are connected to each other with suitably chosen SQUIDs which provide the necessary complex hoppings. The color-entangled Hofstadter model can be realized as shown in panel (c). To get a $C=2$ band, one can put two LCJ circuits (which are represented using red and blue colors respectively) on each lattice sites and connect them to each other according to the Hamiltonian presented in the main text. There are only intra-color hoppings along the $x$ direction within unit cells. The hoppings along the $x$ direction across boundaries between two unit cells are twisted, {\em i.e.}, from red (blue) to blue (red). There are only intra-color hoppings along the $y$ direction. The phase of the blue to blue hopping on a bond is $\pi$ more than the phase of the red to red hopping along the bond. The numbers in panels (b) and (c) are defined in the same way as in Fig.~1 of the main text.}
\label{FigureS3}
\end{figure}

\subsection{Appendix D. Toward Experimental Realization}

In this section, we present more details regarding the two experimental systems in which the bosonic fractional topological phases studied in the main text might be realized. To get bosonic fractional topological phases in the standard Hofstadter model and its variants, there are two essential ingredients that need to be implemented. The first one is complex hopping amplitudes with phases determined by the magnetic field. The second one is that the particles should interact with each other. For the phases of our interest, intra-color onsite interaction is desirable and inter-color onsite interaction should be minimized.

The complex hoppings required for the standard Hofstadter model have been realized in atomic systems using laser-assisted tunneling in optical lattices~\cite{Aidelsburger,Miyake}. To find an experimental protocol for realizing the color-entangled Hofstadter model using cold atoms, we note that in this model a particle may acquire a phase as well as rotate in the internal color space when it hops between two lattice sites. This is a special form of spin-orbit coupling which can in principle be generated if the particles are coupled to suitable non-Abelian gauge fields. By transforming the momentum space Hamiltonian back to real space
\begin{eqnarray}
\Psi^\dagger(mn) e^{iA^{mn}_x} \Psi(m+1,n) + \Psi^\dagger(mn) e^{iA^{mn}_y} \Psi(m,n+1) + {\rm H.c.}
\end{eqnarray}
one can see what are the necessary gauge fields $A^{mn}_{x,y}$. One may extract the Chern number of a system in an optical lattice~\cite{Zhao,Goldman2,Abanin} to demonstrate that a topologically non-trivial model has been realized. It is a good approximation to consider only onsite interaction for bosons in optical lattices. The tunability of interaction in atomic systems~\cite{Zwerger,Chin} may allow us to minimize the inter-color interaction to stabilize the fractional topological phases. One may also devise some methods to directly measure the braiding statistics following the proposals presented in other contexts~\cite{Zhang,Jiang,Aguado}.

We still use the four band $C=2$ model shown in Fig.~\ref{Figure1} as an example. It is easy to check that one would obtain very unnatural gauge fields $A^{mn}_{x,y}$ when using the original Hamiltonian Eq.~(\ref{Hamilton2}), so we want to change it to another form which gives more realistic gauge fields $A^{mn}_{x,y}$. Using gauge transformations $\Psi_{02} \rightarrow U_0\Psi_{02}$ and $\Psi_{13} \rightarrow U_1\Psi_{13}$, where $U_0=\exp[i\phi(1-\sigma_x)]$ and $U_1=\exp[i\theta(1-\sigma_x)]$ induce rotations in the internal color space along the $x$ axis, we can change the Hamiltonian to
\begin{eqnarray}
&& \Psi^\dagger_{02}({\mathbf k}) e^{ik_y} \left(
\begin{array}{cc}
\cos(2\phi) & -i\sin(2\phi) \\
i\sin(2\phi) & -\cos(2\phi) \\
\end{array}
\right) \Psi_{02}({\mathbf k}) +
\Psi^\dagger_{13}({\mathbf k}) e^{i(k_y+\pi/2)} \left(
\begin{array}{cc}
\cos(2\theta) & -i\sin(2\theta) \\
i\sin(2\theta) & -\cos(2\theta) \\
\end{array}
\right) \Psi_{13}({\mathbf k}) \nonumber \\
+ && \Psi^\dagger_{02}({\mathbf k}) e^{ik_x/2} e^{i(\theta-\phi)} \left(
\begin{array}{cc}
\cos(\theta-\phi) & -i\sin(\theta-\phi) \\
-i\sin(\theta-\phi) & \cos(\theta-\phi) \\
\end{array}
\right) \Psi_{13}({\mathbf k}) \nonumber \\
+ && \Psi^\dagger_{13}({\mathbf k}) e^{ik_x/2} e^{i(\pi/2-\theta+\phi)} \left(
\begin{array}{cc}
\cos(\pi/2-\theta+\phi) & -i\sin(\pi/2-\theta+\phi) \\
-i\sin(\pi/2-\theta+\phi) & \cos(\pi/2-\theta+\phi) \\
\end{array}
\right) \Psi_{02}({\mathbf k}) + {\rm H.c.}
\end{eqnarray}
By choosing $\phi=0$ and $\theta=\pi/4$, we get
\begin{eqnarray}
&& \Psi^\dagger_{02}({\mathbf k}) e^{ik_y} \left(
\begin{array}{cc}
1 & 0 \\
0 & -1 \\
\end{array}
\right) \Psi_{02}({\mathbf k}) +
\Psi^\dagger_{13}({\mathbf k}) e^{i(k_y+\pi/2)} \left(
\begin{array}{cc}
0 & -i \\
i & 0 \\
\end{array}
\right) \Psi_{13}({\mathbf k}) \nonumber \\
+ && \Psi^\dagger_{02}({\mathbf k}) e^{i(k_x/2+\pi/4)} \left(
\begin{array}{cc}
\cos(\pi/4) & -i\sin(\pi/4) \\
-i\sin(\pi/4) & \cos(\pi/4) \\
\end{array}
\right) \Psi_{13}({\mathbf k}) \nonumber \\
+ && \Psi^\dagger_{13}({\mathbf k}) e^{i(k_x/2+\pi/4)} \left(
\begin{array}{cc}
\cos(\pi/4) & -i\sin(\pi/4) \\
-i\sin(\pi/4) & \cos(\pi/4) \\
\end{array}
\right) \Psi_{02}({\mathbf k}) + {\rm H.c.}
\end{eqnarray}
which means that
\begin{eqnarray}
A^{mn}_x = e^{i\pi/4} \left(
\begin{array}{cc}
\cos(\pi/4) & -i\sin(\pi/4) \\
-i\sin(\pi/4) & -\cos(\pi/4) \\
\end{array}
\right) \;\;\;
A^{mn}_y = e^{im\pi/2}\left(
\begin{array}{cc}
\cos(m\pi/2) & -i\sin(m\pi/2) \\
i\sin(m\pi/2) & -\cos(m\pi/2) \\
\end{array}
\right)
\end{eqnarray}
If this method were to be realized in experiments, we need to facilitate rotations in the internal color space when a particle hops in both the $x$ and $y$ directions. It can be simplified if we start from a slightly different Hamiltonian
\begin{eqnarray}
&& \Psi^\dagger_{02}({\mathbf k}) e^{ik_y} \left(
\begin{array}{cc}
0 & 1 \\
1 & 0 \\
\end{array}
\right) \Psi_{02}({\mathbf k}) +
\Psi^\dagger_{13}({\mathbf k}) e^{i(k_y+\pi/2)} \left(
\begin{array}{cc}
0 & 1 \\
1 & 0 \\
\end{array}
\right) \Psi_{13}({\mathbf k}) \nonumber \\
+ && \Psi^\dagger_{02}({\mathbf k}) e^{ik_x/2} \left(
\begin{array}{cc}
1 & 0 \\
0 & 1 \\
\end{array}
\right) \Psi_{13}({\mathbf k}) +
\Psi^\dagger_{13}({\mathbf k}) e^{ik_x/2} \left(
\begin{array}{cc}
1 & 0 \\
0 & -1 \\
\end{array}
\right) \Psi_{02}({\mathbf k}) + {\rm H.c.}
\end{eqnarray}
This Hamiltonian is equivalent to the color-entangled Bloch basis if the $P$ and $Q$ operators are exchanged in the construction. Using gauge transformations $U_0=\exp(-i\phi\sigma_y)$ and $U_1=\exp(-i\theta\sigma_y)$, this Hamiltonian is changed to
\begin{eqnarray}
&& \Psi^\dagger_{02}({\mathbf k}) e^{ik_y} \left(
\begin{array}{cc}
0 & e^{i\phi} \\
e^{-i\phi} & 0 \\
\end{array}
\right) \Psi_{02}({\mathbf k}) +
\Psi^\dagger_{13}({\mathbf k}) e^{i(k_y+\pi/2)} \left(
\begin{array}{cc}
0 & e^{i\theta} \\
e^{-i\theta} & 0 \\
\end{array}
\right) \Psi_{13}({\mathbf k}) \nonumber \\
+ && \Psi^\dagger_{02}({\mathbf k}) e^{ik_x/2} \left(
\begin{array}{cc}
1 & 0 \\
0 & e^{i(\theta-\phi)} \\
\end{array}
\right) \Psi_{13}({\mathbf k}) +
\Psi^\dagger_{13}({\mathbf k}) e^{ik_x/2} \left(
\begin{array}{cc}
1 & 0 \\
0 & e^{i(\pi-\theta+\phi)} \\
\end{array}
\right) \Psi_{02}({\mathbf k}) + {\rm H.c.}
\end{eqnarray}
By choosing $\phi=0$ and $\theta=\pi/2$, we have
\begin{eqnarray}
A^{mn}_x = \left(
\begin{array}{cc}
0 & e^{im\pi} \\
1 & 0 \\
\end{array}
\right) \;\;\; 
A^{mn}_y = 
\left(
\begin{array}{cc}
1 & 0 \\
0 & i \\
\end{array}
\right)
\end{eqnarray}
and a particle only rotates in the internal color space when it hops along the $x$ direction.

We now discuss how the fractional topological states of our interest might be generated using photons. It was proposed in Ref.~\cite{Hafezi2} that the standard Hofstadter model can be realized in a circuit quantum electrodynamics system and the photonic excitations in such a system may form fractional quantum Hall states. The proposal we present here is largely motivated by this work so we first explain the system proposed in Ref.~\cite{Hafezi2} as shown in panel (b) of Fig.~\ref{FigureS3}. There is one set of LCJ circuit made of one inductor, one capacitor, and one Josephson junction on each lattice site. The LCJ circuit can be described using the following Hamiltonian~\cite{Devoret}
\begin{eqnarray}
H_{\rm LCJ} = \frac{\phi^2}{2L} + \frac{Q^2}{2C} - E_J \cos \left( \frac{\phi+\alpha}{\phi_s} \right)
\end{eqnarray}
where $\phi$ is the node flux, $L$ is the inductance, $Q$ is the node charge, $C$ is the capacitance, $E_J$ is the Josephson energy, $\alpha$ is the external flux through the Josephson junction, and $\phi_s=\hbar/2e$ is the reduced flux quantum. The three terms in this Hamiltonian represent the shunted inductive energy, the electron charging energy, and the Josephson energy with characteristic scales $E_L=\phi^2_s/2L$, $E_c=e^2/2C$, and $E_J$, respectively. In the transmon regime with $E_L \sim E_J \gg E_c$~\cite{Koch2}, the Hamiltonian can be reduced to 
\begin{eqnarray}
H_{\rm LCJ} = \hbar \omega a^\dagger a + V_2 a^\dagger a^\dagger a a + {\rm higher \; order \; terms}
\end{eqnarray}
where $a$ is the annihilation operator for photonic excitations in the circuit, $\omega$ is the resonant frequency determining the single-particle energy scale, and $V_2$ characterizes the strength of the two-body photon-photon interaction. The higher order terms contain three-body and other multi-particle interactions between photons, which may be exploited for realizing some fractional topological states but are not needed for our discussion here. The frequencies on the lattice sites should be set appropriately to form a staggered pattern. The LCJ circuits are coupled to each other using superconducting quantum interference devices (SQUIDs) to implement complex hoppings between the lattice sites. The SQUID between the lattice sites labeled by $i$ and $j$ has an inductance that can be controlled using an external microwave flux $\phi(t)=\delta\phi\cos(\Delta_{ij} t + \theta_{ij})$. The coefficient $\delta\phi$ is chosen to be much smaller than $1$ and the microwave pump frequency $\Delta_{ij}$ is tuned to be the same as the difference between the resonant frequencies $\omega_i$ and $\omega_j$. When one changes to the rotating frame and adopt the rotating wave approximation, the lattice sites labeled by $i$ and $j$ is connected by the hopping term $a^{\dagger}_i a_j e^{i\theta_{ij}}$+{\rm H.c.}.

The advantage of this approach is that two lattice sites can be connected on demand and the hopping phase between them can be tuned using the SQUID. This motivates us to construct the system shown in panel (c) of Fig.~\ref{FigureS3} to realize the color-entangled Hofstadter model with $C=2$ band. There are two sets of LCJ circuit on each lattice sites (which we label as red and blue as in the main text) and they are connected to their neighbors as required by the Hamiltonian of the color-entangled Hofstadter model. For the hoppings along the $y$ direction on any bond, the phase between two blue circuits is $\pi$ more than the phase between two red circuits. It is also possible to apply regular hoppings [from red (blue) sites to red (blue) sites] within unit cells and twisted hoppings [from red (blue) sites to blue (red) sites] at the boundaries between two unit cells. There only exist interaction between two photons that are in the same LCJ circuit so the many-body Hamiltonian have only intra-color onsite interaction, which is desirable for stabilizing the bosonic fractional topological phases of interest. 

\end{widetext}

\end{appendix}

\end{document}